\documentclass[runningheads,a4paper]{llncs}

\usepackage{csquotes} 

\usepackage{graphicx}

\usepackage{microtype}

\usepackage{listings}

\usepackage{subfigure}

\usepackage{tabularx}
\usepackage{multirow}

\usepackage{pgfplots,wrapfig}

\makeatletter
\newcommand\footnoteref[1]{\protected@xdef\@thefnmark{\ref{#1}}\@footnotemark}
\makeatother

\usepackage{tikz}
\usetikzlibrary{matrix,positioning}

\definecolor{redi}{RGB}{255,38,0}
\definecolor{redii}{RGB}{200,50,30}
\definecolor{yellowi}{RGB}{255,251,0}
\definecolor{bluei}{RGB}{0,150,255}
\definecolor{blueii}{RGB}{135,247,210}
\definecolor{blueiii}{RGB}{91,205,250}
\definecolor{blueiv}{RGB}{115,244,253}
\definecolor{bluev}{RGB}{1,58,215}
\definecolor{orangei}{RGB}{240,143,50}
\definecolor{yellowii}{RGB}{222,247,100}
\definecolor{greeni}{RGB}{166,247,166}

\tikzset{ 
	table/.style={
		matrix of nodes,
		row sep=-\pgflinewidth,
		column sep=-\pgflinewidth,
		nodes={rectangle,draw=black,text width=1.25ex,align=center},
		text depth=0.25ex,
		text height=1ex,
		nodes in empty cells
	},
	texto/.style={font=\footnotesize\sffamily},
	title/.style={font=\small\sffamily}
}

\newcommand\CellText[2]{%
	\node[texto,left=of mat#1,anchor=east]
	at (mat#1.west)
	{#2};
}

\newcommand\SlText[2]{%
	\node[texto,left=of mat#1,anchor=west,rotate=75]
	at ([xshift=3ex]mat#1.north)
	{#2};
}

\newcommand\RowTitle[2]{%
	\node[title,left=6.3cm of mat#1,anchor=west]
	at (mat#1.north west)
	{#2};
}

\usepackage{url}
\urldef{\mailsa}\path|daniel.ritter@sap.com |
\newcommand{\keywords}[1]{\par\addvspace\baselineskip
	\noindent\keywordname\enspace\ignorespaces#1}

\newcommand{\eg}{e.\,g., }
\newcommand{\ie}{i.\,e., }

\begin{document}
	
	\mainmatter  
	
	\title{ 
		Towards More Data-Aware Application Integration (extended version)}
	
	\titlerunning{ 
		Towards More Data-Aware Application Integration (extended version)}
	
	%
	%
	\author{Daniel Ritter}
	\authorrunning{Daniel Ritter}
	
	\institute{SAP SE, Technology Development,\\
		Dietmar-Hopp-Allee 16, 69190 Walldorf, Germany\\
		\mailsa\\
	}
	
	%
	%
	
\tocauthor{Authors' Instructions}
\maketitle
	
\begin{abstract}
Although most business application data is stored in relational databases, programming languages and wire formats in integration middleware systems 
are not table-centric. Due to costly format conversions, data-shipments and faster computation, the trend is to \enquote{push-down} the integration operations closer to the storage representation.
We address the alternative case of defining declarative, table-centric integration semantics within standard integration systems. For that, we replace the current operator implementations for the well-known \emph{Enterprise Integration Patterns} by equivalent \enquote{in-memory} table processing, and show a practical realization in a conventional integration system for a non-reliable, \enquote{data-intensive} messaging example. The results of the runtime analysis show that table-centric processing is promising already in standard, \enquote{single-record}  message routing and transformations, and can potentially excel the message throughput for \enquote{multi-record} table messages.
\end{abstract}
	
\keywords{Datalog, Message-based / Data integration, Integration System.}
	
\section{Introduction} \label{sec:introduction}
Integration middleware systems in the sense of EAI brokers \cite{Chappell:2004:ESB:1237904}
(\eg SAP HANA Cloud Integration\footnote{SAP HCI, visited 04/2015: \url{https://help.sap.com/cloudintegration}.}, Boomi AtomSphere\footnote{Boomi AtomSphere, visited 04/2015: \url{http://www.boomi.com/integration}.})
address the fundamental need for (business) application integration by acting as a messaging hub between applications. As such, they have become ubiquitous in service-oriented enterprise computing environments. 
Messages are mediated between applications mostly in wire formats based on XML (\eg SOAP for Web Services).
	
The advent of more \enquote{data-aware} integration scenarios (observation \emph{O1}) put emphasis on (near) \enquote{real-time} or online processing (\emph{O2}), which requires us to revisit the standard integration capabilities, system design and architectural decisions. For instance, in the financial / utilities industry, \emph{China Mobile} generates 5-8 TB of call detail records per day, which have to be processed by integration systems (\ie mostly message routing and transformation patterns), \enquote{convergent charging}\footnote{\label{foot:solace}Solace Solutions, visited 02/2015; last update 2012: \url{http://www.solacesystems.com/techblog/deconstructing-kafka}.} (CC) and \enquote{invoicing} applications (not further discussed). In addition, the standard XML-processing has to give ground to other formats like JSON and CSV (\emph{O3}). These observations (\emph{O1--3}) are backed by similar scenarios from sports management (\eg online player tracking) and the rapidly growing amount of data from the \emph{Internet of Things} and \emph{Cyber Physical System} domains. For those scenarios, an architectural setup with systems like \emph{Message Queuing} (MQ) are used as reliable \enquote{message buffers} (\ie queues, topics) that handle \enquote{bursty} incoming messages and smoothen peak loads (cf. Fig. \ref{fig:approach}). Integration systems are used as message consumers, which (transactionally) dequeue, transform (\eg mapping, content enrichment) and route messages to applications. For reliable transport and message provenance, integration systems require relational Database Systems, in which most of the (business) application data is currently stored (\emph{O4}). When looking at the throughput capabilities of the named systems, software-/hardware-based MQ systems like \emph{Apache Kafka} or \emph{Solace}\footnoteref{foot:solace} are able to process several millions of messages per second. RDBMS benchmarks like TPC-H measure queries and inserts in PB sizes, while simple, header-based routing benchmarks for integration systems show message throuphputs of few thousands of messages per second \cite{esbperformance2013} (\emph{O5}). In other words, MQ and DBMS (\eg RDBMS, NoSQL, NewSQL) systems are already addressing observations \emph{O1--5}. Integration systems, however, seem to not be there yet.

Compared to MQs, integration systems work on message data, which seems to make the difference in message throughput. We argue that integration operations, represented by  \emph{Enterprise Integration Patterns} (EIP) \cite{Hohpe:2003:EIP:940308}, can be mapped to an \enquote{in-memory} representation of the table-centric RDBMS operations to profit from their efficient and fast evaluation. Early ideas on this were brought up in our position papers \cite{DBLP:conf/zeus/Ritter14,DBLP:conf/dexa/RitterB14}. In this work, we follow up to shed light on the observed discrepancies. We revisit the EIP building blocks and operator model of integration systems, for which we define RDBMS-like table operators (so far without changing their semantics) as a symbiosis of RDBMS and integration processing by using Datalog \cite{DBLP:books/cs/Ullman88}. We choose Datalog as example of an efficiently computable, table-like integration co-processing facility close to the actual storage representation with expressive foundations (\eg recursion), which we call \emph{Table-centric Integration Patterns} (TIP). To show the applicability of our approach to integration scenarios along observations \emph{O1--5} we conduct an experimental message throughput analysis for selected routing and transformation patterns, where we carefully embed the TIP definitions into the open-source integration system \emph{Apache Camel} \cite{apacheCamel13} that implements most of the EIPs. Not changing the EIP semantics means that table operations are executed on \enquote{single-record} table messages. We give an outlook to \enquote{multi-record} table message processing.
	
The remainder of this paper is organized along its main contributions. After a more comprehensive explanation of the motivating CC example and a brief sketch of our approach in Sect. \ref{sec:approach}, we analyse common integration patterns with respect to their extensibility for alternative operator models and define a table-centric operator / processing model that can be embedded into the patterns (still) aligned with their semantics using for example Datalog in Sect. \ref{sec:datapatterns}. In Sect. \ref{sec:eval} we apply our approach to a conventional integration system and briefly describe and discuss our experimental performance analysis, and we draw an experimental sketch of the idea of \enquote{multi-record} table message processing. Section \ref{sec:relatedwork} examines related work and Sect. \ref{sec:discussion} concludes the paper.

\section{Motivating Example and General Approach} \label{sec:approach}
\begin{figure}[!ht]
	\centering
	\includegraphics[width=0.9\columnwidth]{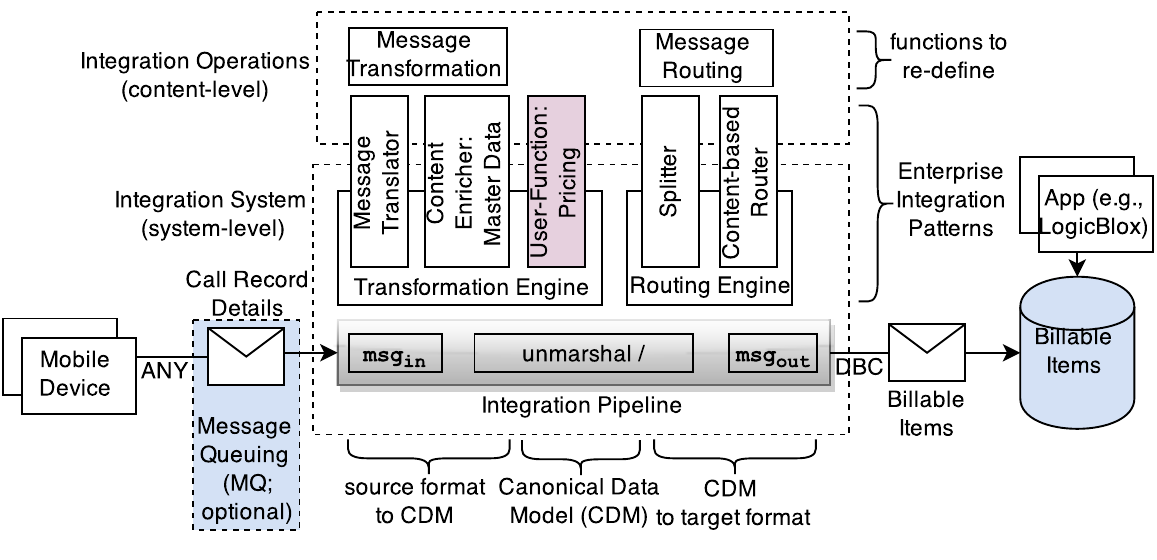}
	\caption{High-level overview of the convergent charging application and architecture.}
	\label{fig:approach}
\end{figure}

In this section, the motivating \enquote{Call Record Detail} example in the context of the \enquote{Convergent Charging} application is described more comprehensively along with a sketch of our approach. Figure \ref{fig:approach} shows aspects of both as part of a common integration system architecture.

\subsection{The Convergent Charging Scenario}
Mobile service providers like \emph{China Mobile} generate large amounts of \enquote{Call Record Details} (CRDs) per day that have been processed by applications like SAP Convergent Charging\footnote{SAP Convergent Charging, last visited 04/2015: \url{https://help.sap.com/cc}.} (CC). As shown in Fig. \ref{fig:approach}, these CRDs are usually sent from mobile devices to integration systems (optionally buffered in a MQ System), where they are translated to an intermediate (application) format and enriched with additional master data (\eg business partner, product). The master data helps to calculate pricing information, with which the message is split into several messages, denoting billable items (\ie item for billing) that are routed to their receivers (\eg DB). From there applications like \emph{SAP Convergent Invoicing} generate legally binding payment documents. Alternatively, new application and data analytics stacks like \emph{LogicBlox} \cite{DBLP:conf/datalog/GreenAK12}, \emph{WebdamLog} \cite{Abiteboul:2013:RAD:2463676.2465251}, and \emph{SAP S/4HANA}\footnote{SAP S/4HANA, last visited 04/2015: \url{http://discover.sap.com/S4HANA}} (not shown) access the data for further processing. Some of these \enquote{smart} stacks even provide declarative, Datalog-like language for application and user-interface programming, which complements our integration approach. As motivated before, standard integration systems have problems processing the high number and rate of incoming messages, which usually leads to an \enquote{offline}, multi-step processing using indirections like ETL systems and pushing integration logic to the applications, leading to long-running CC runs. 

\subsection{General Approach}
The \emph{Enterprise Integration Patterns} (EIPs) \cite{Hohpe:2003:EIP:940308} define 
\enquote{de-facto} standard operations on the header (\ie payload's meta-data) and body (\ie message payload) of a message, which are normally implemented in the integration system's host language (\eg Java, C\#). This way, the actual integration operation (\ie the content developed by an integration expert like mapping programs and routing conditions) can be differentiated from the implementation of the runtime system that invokes the content operations and processes their results. For instance, Fig. \ref{fig:approach} shows the separation of concerns within integration systems with respect to \enquote{system-related} and \enquote{content-related parts} and sketches which pattern operations to re-define using relational table operators, while leaving the runtime system (implementation) as is. The goal is to only change these operations and make integration language additions for table-centric processing within the conventional integration system, while preserving the general integration semantics like \emph{Quality of Service} (\eg best effort, exactly once) and the \emph{Message Exchange Pattern} (\eg one-way, two-way). In other words, the content-related parts of the pattern definitions are evaluated by an \enquote{in-process} table operation processor (\eg a Datalog system), which is embedded into the standard integration system and invoked during the message processing.
	
\section{Table-centric Integration Patterns} \label{sec:datapatterns}
Before defining \emph{Table-centric Integration Patterns} (short \texttt{TIP}) for message routing and transformation more formally, let us recall the encoding of some
relevant, basic database operations / operators into Datalog: \texttt{join}, \texttt{projection}, \texttt{union}, and \texttt{selection}. The join of two relations $r(x, y)$ and $s(y, z)$ on parameter $y$ is encoded as $j(x, y, z) \leftarrow r(x, y), s(y, z)$, which projects all three parameters to the resulting predicate $j$. More explicitly, a projection on parameter $x$ of relation $r(x, y)$ is encoded as $p(x) \leftarrow r(x, y)$. The union of $r(x, y)$ and $s(x, y)$ is $u(x, y) \leftarrow r(x, y).$ $u(x, y) \leftarrow s(x, y)$, which combines several relations to one. The selection $r(x, y)$ according to a built-in predicate $\phi(x)$, where $\phi(x)$ can contain constants and free variables, is encoded as $s(x, y) \leftarrow r(x, y), \phi(x)$.
Built-in predicates can be binary relations on numbers such as $<$,$<=$,$=$, binary relations on strings such as $equals, contains, starts with$ or predicates applied to expressions based on binary operators like $+,-,*,/$ (\eg $x=p(y)+1$), and operations on relations like $z=max(p(x,y), x),z=min(p(x,y),x)$, which would assign the maximal or the minimal value $x$ of a predicate $p$ to a parameter $z$.

Although our approach allows each single pattern definition to evaluate arbitrary, recursive Datalog operations and built-in predicates, the Datalog to pattern mapping tries to identify and focus on the most relevant table-centric operations for a specific pattern. An overview of the mapping of all discussed message routing and transformation operations to Datalog constructs is shown in Fig. \ref{fig:routetransform} and is subsequently discussed.
\begin{figure}
	\begin{tikzpicture}[node distance =0pt and 0.5cm]
	\matrix[table] (mat11) 
	{
		|[fill=blue]| & |[fill=bluei]| & |[fill=blue]| & & \\
		|| & & & |[fill=blue]| & \\
		|| & & & & \\
		|[fill=bluei]| & |[fill=bluei]| & |[fill=blue]| & & \\
		|[fill=blue]| & & |[fill=blue]| & & \\
		|| & & & & |[fill=blue]| \\
	};
	
	\matrix[table,below=of mat11] (mat21) 
	{
		|[fill=bluei]| & |[fill=blue]| & & |[fill=blue]| & \\
		|[fill=blue]| & & |[fill=blue]| & |[fill=blue]| & \\
		|| & & & & |[fill=blue]| \\
	};
	
	\SlText{11-1-1}{built-in}
	\SlText{11-1-2}{join}
	\SlText{11-1-3}{selection}
	\SlText{11-1-4}{projection}
	\SlText{11-1-5}{union}
	
	\RowTitle{11}{Message Routing};
	\CellText{11-1-1}{Router, Filter: };
	\CellText{11-2-1}{Recipient List};
	\CellText{11-3-1}{Multicast, Join Router};
	\CellText{11-4-1}{Splitter};
	\CellText{11-5-1}{Correlation, Completion};
	\CellText{11-6-1}{Aggregation};
	
	\RowTitle{21}{Message Transformation};
	\CellText{21-1-1}{Message translator};
	\CellText{21-2-1}{Content filter};
	\CellText{21-3-1}{Content enricher};
	\end{tikzpicture}
	\caption{Message routing and transformation patterns mapped to Datalog. Most common Datalog operations for a single pattern are marked \enquote{dark blue}, less common ones \enquote{light blue}, and possible but uncommon ones \enquote{white}.} \label{fig:routetransform}
\end{figure} 
Subsequently, we enumerate common EIPs and separate system- from content-related parts more formally for the TIP definition by example of standard Datalog.
	
\subsection{Canonical Data Model}
When connecting applications, various operations are executed on the transferred messages in a uniform way. The arriving message instances are converted into an internal format understood by the pattern implementation, called the \emph{Canonical Data Model} (CDM) \cite{Hohpe:2003:EIP:940308}, before the messages are transformed to the target format. Hence, if a new application is added to the integration solution, only conversions between the CDM and the application format have to be created. Consequently, for a table-centric re-definition of integration patterns, we define a CDM similar to relational database tables as \emph{Datalog programs}, which consists of a collection of facts / a table, optional (supporting) rules as message body and an optional set of meta-facts that describes the actual data as header. For instance, the data-part of an incoming message in JSON format is transformed to a collection of \texttt{Open-Next-Close} (ONC)-style table iterators, each representing a table row or fact. These ONC-operators are part of the evaluated execution plan for more efficient evaluation.
	
\subsection{Message Routing Patterns}
In this section the message routing pattern implementations are re-defined, which can be seen as control and data flow definitions of an integration channel pipeline. For that, they access the message to route it within the integration system and eventually to its receiver(s). They influence the channel and message cardinality as well as the content of the message.

\paragraph{Content-based Router / Message Filter}The most common routing patterns that determine the message's route based on its body are the \emph{Content-based Router} and the \emph{Message Filter}. The stateless router has a channel cardinality of $1$:$n$, where $n$ is the number of leaving channels, while one channel enters the router, and a message cardinality of $1$:$1$. The entering message constitutes the leaving message according to the evaluation of a \emph{routing condition}. This condition is a function $rc$, with $\{bool_1, bool_2, ..., bool_n\} := rc(msg_{in},conds)$, where $msg_{in}$ is the entering message. The function $rc$ evaluates to a list of Boolean output $\{bool_1, bool_2, ..., bool_n\}$ based on a list of conditions $conds$ of the same arity (\eg Datalog rules in Suppl. Material, List. \ref{lst:cbr}) for each of the $n \in N$ leaving channels. In case several conditions evaluate to \texttt{true}, only the first matching channel receives the message.

Through the separation of concerns, a system-level routing function provides the entering message $msg_{in}$ to the content-level implementation (\ie in CDM representation), which is configured by $conds$. Since standard Datalog rules are truth judgements, and hence do not directly produce Boolean values, 
we decided, for performance and generality considerations, to add an additional function $bool_{rc}$ to the integration system. The function $bool_{rc}$ converts the output list $fact$ of the routing function from a truth judgement to a Boolean by emitting \texttt{true} if $fact \neq \emptyset$, and \texttt{false} otherwise. Accordingly we define the TIP routing condition as $fact := rc_{tip}(msg_{in},conds)$, while being evaluated for each channel condition (\eg selection / built-in predicates). The integration system will then use the function $bool_{rc}$ to convert this into a Boolean value. For the message filter, which is a special case of the router that differs only from its channel cardinality of $1$:$1$ and message cardinality of $1$:$[0|1]$, the filter condition is equal to $rc_{tip}$.


\paragraph{Multicast / Recipient List / Join Router}The stateless \emph{Multicast} and \emph{Recipient List} patterns route multiple messages to several leaving channels, which gives them a message and channel cardinality of $1$:$n$. While the multicast statically routes messages to the leaving channels (\ie no re-definition required), the recipient list determines the channels dynamically. The receiver determination function $rd$, with $\{out_1, out_2, ..., out_n\} := rd(msg_{in}, [header.y|body.x]).$, computes $n \in N$ receiver channel configurations $\{recv_1, recv_2, ..., recv_n\}$ by extracting their key values either from an arbitrary message header field $header.y$ or from a message body field $body.x$. The integration system has to implement a receiver determination function that takes the list of key-strings $\{recvId_1, recvId_2, ..., recvId_m\}$ as input, for which it looks up receiver configurations $recv_{0}, recv_{1}, ..., recv_{n}$, where $m,n \in N$ and $m > n$, and routes copies of the entering message $\{msg_{out}', msg_{out}'', ..., msg_{out}^{n'}\}$.

In terms of TIP, $rd_{tip}$ is a projection of message body or header values to a unary, output relation. For instance, the receiver configuration keys $recvId_1$ and $recvId_2$ have to be part of the message body like $body(x,'recvId_1').\allowbreak body(x,'recvId_2').$. Then the $rd_{tip}$ would evaluate a Datalog rule similar to $config(y) \leftarrow body(x, y)$, while the keys $recvId_1$ and $recvId_2$ correspond to receiver configurations $\{recv_1, recv_2\}$. 

\paragraph{Splitter / Aggregator}The antipodal \emph{Splitter} and \emph{Aggregator} patterns both have a channel cardinality of $1$:$1$ and create new, leaving messages. Thereby the splitter breaks the entering message into multiple (smaller) messages (\ie message cardinality of $1$:$n$) and the aggregator combines multiple entering messages to one leaving message (\ie message cardinality of $n$:$1$).
Hereby, the stateless splitter uses a split condition $sc$ on the content-level, with $\{out_1, out_2,..., out_n\} := sc(msg_{in},conds)$, which accesses the entering message's body to determine a list of distinct body parts $\{out_1, out_2, ..., out_n\}$, based on a list of conditions $conds$, that are each inserted to a list of individual, newly created, leaving messages $\{msg_{out1}, msg_{out2}, ..., msg_{outn}\}$ with $n \in N$ by a splitter function. The header and attachments are copied from the entering to each leaving message.

The re-defined split condition $sc_{tip}$ evaluates a set of Datalog rules as $conds$ (\ie mostly selection, and sometimes built-in and join constructs; the latter two are marked \enquote{light blue}). Each part of the body $out_i$ with $i \in N$ is a set of facts that is passed to a split function, which wraps each set into a single message.

The stateful aggregator defines a correlation condition, completion condition and an aggregation strategy. The correlation condition $crc$, with $coll_i := crc(msg_{in}, \allowbreak conds)$, determines the aggregate collection $coll_i$, to which the message is stored, based on a list of conditions $conds$. The completion condition $cpc$, with $cp_{out} := cpc(msg_{in},[header.y|body.x])$, evaluates to a Boolean output $cp_{out}$ based on header or body field information (similar to the message filter). If $cp_{out}$ equals \texttt{true}, then the aggregation strategy $as$, with $agg_{out} := as({msg^1_{in}, msg^2_{in}, ..., msg^n_{in}})$, is called by an implementation of the messaging system and executed, else the current message is added to the collection $coll_i$. The $as$ evaluates the correlated entering messages $coll_i$ and emits a new message $msg_{out}$. For that, the messaging system has to implement an aggregation function that takes $agg_{out}$ (\ie the output of $as$) as input.

These functions are re-defined as $crc_{tip}$ and $cpc_{tip}$ such that the $conds$ are Datalog rules mainly with selection and built-in constructs. The $cpc_{tip}$ makes use of the defined $bool_{rc}$ function to map its evaluation result (\ie list of facts or empty) to the Boolean value $cpout$. The aggregation strategy $as$ is re-defined as $as_{tip}$, which mainly uses \texttt{union} to combine lists of facts from different messages to one. The message format remains the same. To transform the aggregates' formats, a message translator is used to keep the patterns modular.
However, the combination of the aggregation strategy with translation capabilities could lead to runtime optimizations. 

\subsection{Message Transformation Patterns}
The transformation patterns exclusively target the content of the messages in terms of format conversations and content modifications.

The stateless \emph{Message Translator} changes the structure or format of the entering message without generating a new one (\ie channel, message cardinality $1$:$1$). For that, the translator computes the transformed structure by evaluating a mapping program $mt$ (\eg Datalog rules in Suppl. Material, List. \ref{lst:mapping}), with $msg_{out}.body := mt(msg_{in}.body)$. Thereby the field content can be altered. The related \emph{Content Filter} and \emph{Content Enricher} patterns can be subsumed by the general \emph{Content Modifier} pattern and share the same characteristics as the translator pattern. The filter evaluates a filter function $mt$, which only filters out parts of the message structure (\eg fields or values) and the enricher adds new fields or values as $data$ to the existing content structure using an enricher program $ep$, with $msg_{out}.body := ep(msg_{in}.body,data)$.

The re-definition of the transformation function $mt_{tip}$ for the message translator mainly uses \texttt{join} and \texttt{projection} (plus \texttt{built-in} for numerical calculations and string operations, thus marked \enquote{light blue}) and \texttt{selection}, \texttt{projection} and \texttt{built-in} (mainly numerical expressions and character operations) for the content filter. While projections allow for rather static, structural filtering, the built-in and selection operators can be used to filter more dynamically based on the content. The resulting Datalog programs are passed as $msg_{out}.body$. In addition, the re-defined enricher program $ep_{tip}$ mainly uses \texttt{union} operations to add additional $data$ to the message as Datalog programs. 

\subsection{Pattern Composition} Since the TIP definitions target the content-level, all patterns can still be composed to more complex integration programs (\ie integration scenarios or pipelines). From the many combinations of patterns, we briefly discuss two important structural patterns that are frequently used in integration scenarios: (1) scatter/gather and (2) splitter/gather \cite{Hohpe:2003:EIP:940308}. The scatter/gather pattern (with a $1$:$n$:$1$ channel cardinality) is a multicast or recipient list that copies messages to several, statically or dynamically determined pipeline configurations, which each evaluate a sequence of patterns on the messages in parallel. Through an aggregator pattern, the messages are structurally and content-wise joined. The splitter/gather pattern (with a $1$:$n$:$1$ message cardinality) splits one message into multiple parts, which can be processed in parallel. In contrast to the scatter/gather the pattern sequence is the same for each instance. A subsequently configured aggregator combines the messages to one.
	
\section{Experimental Evaluation} \label{sec:eval}
As \emph{System under Test} (SuT) for an experimental evaluation we used the open source, Java-based \emph{Apache Camel} integration system \cite{apacheCamel13} in version 2.14.0, which implements most of the EIPs. The Camel system allows content-level extensions through several interfaces, with which the TIP definitions were implemented and embedded (\eg own Camel \texttt{Expression} definitions for existing patterns, and Camel \texttt{Processor} definitions for custom or non-supported patterns). The Datalog system we used for the measurements is a Java-based, standard na\"{i}ve-recursive Datalog processor (\ie without stratification) \cite{DBLP:books/cs/Ullman88} in version 0.0.6 from \cite{DBLP:conf/datalog/RitterW12}.

Subsequently, the basic setup and execution of the measurements are introduced. However, due to brevity, a more detailed description of the setup is provided in the Suppl. Material, Sect. \ref{supp:setupDetails}, the routing condition and mapping programs are shown in Sect. \ref{app:A}, the integration scenarios in Sect. \ref{supp:routes} and the more detailed results in Sect. \ref{supp:results}.
	
\subsection{Setup}
In the absence of an EIP benchmark, which we are currently developing on the basis of this paper, we used \emph{Apache JMeter}\footnote{Apache JMeter, visited 02/2015: \url{http://jmeter.apache.org/}.} in version 2.12 as a load generator client that sends messages to the SuT. We implemented a JMeter Sampler, which allows to inject messages directly to the integration pipeline via a Camel \texttt{direct} endpoint / adapter. For the throughput measurements, we used the JMeter \texttt{jp@gc} transaction per second listener plugin from the standard package.

To measure the message throughput in a \enquote{data-intensive} (cf.\ \emph{O1}), non-reliable integration scenario, we use the standard TPC-H order, customer and nation data sets. We added additional, unique message identifier and type fields and translate the single records to JSON objects (cf.\ \emph{O3}), each representing the payload of a single message (\ie \enquote{single-record} table message). In this way we generated $1.5$ million order-only messages (\ie TPC-H scale level 1) and the same amount of \enquote{multi-format} customer / nation messages, consisting of one customer and all $25$ nation records per message (in the \enquote{single-record} table message case). During the  measurements these messages are streamed to the Camel endpoint, serialized to either Java Objects for the Camel-Java and to the ONC representation for the Camel-Datalog case (cf. recall ONC-iterators as canonical data model).
	
All measurements are conducted on a HP Z600 work station, equipped with two Intel processors clocked at 2.67GHz with a total of 12 cores, 24GB of main memory, running a 64-bit Windows 7 SP1 and a JDK version 1.7.0. The JMeter Sampler and the integration system pipeline JVM process get $5$GB heap space.
	
\subsection{\enquote{Single-record} / \enquote{Multi-Format} Table Message Processing}
Instead of testing all discussed patterns, we focus on the identified table-operations (\eg \texttt{selection} / \texttt{built-in}, \texttt{projection}, \texttt{join}) and show the respective evaluation by example of a representative pattern (cf. Fig. \ref{fig:routetransform}). The measurements for selection and projection use the TPC-H \texttt{Order}-based, approximately $4$kB messages (\ie $1.5$ million order messages). The \texttt{union} operation (\eg aggregation strategy, content enricher) is not tested.
	
We measured the selection / built-in operations in a content-based router scenario with a routing condition $tip\_{rc}$ (cf. List. \ref{lst:cbr}), which routes the order message to its receiver based on \emph{conds} for  \{string equality, integer less than\} on fields \{objecttype, ototalprice\}. The $bool_{rc}$ function is implemented in Java to pass the expected value to the runtime system on system-level. The corresponding \enquote{hand-coded} content-level Camel-Java implementation uses JSON path statements for $O(1)$ element access and conducts the type-specific condition evaluation. The routing condition is defined to route $904,500$ of the $1.5$million messages to the first and the rest to the second receiver. Similarly, the projection operation is measured using a message translator. The translator projects the fields of the incoming order message to a target format (cf. List. \ref{lst:mapping}) using a $mt_{tip}$ implementation or a \enquote{hand-coded} projection on the Java Object representation. Now, the \enquote{multi-format} customer messages (cf. \emph{O1}) with nation records as processing context are used to measure a routing condition with selection / built-in and join operations (cf. List. \ref{lst:join}). The customer message is routed, if and only if, the customers balance (\texttt{ACCTBAL}) is bigger than $3,000$ and the customer is from the European region determined through join via nation key.
\par\noindent
\begin{minipage}[t]{.45\textwidth}
	\begin{lstlisting}[language=Java,caption=Routing condition: $tip\_{rc}$,label={lst:cbr},numbers=left,numbersep=1pt]
cbr-order(id,-,OTOTALPRICE,-):-
order(id,otype,-,
OTOTALPRICE,-OPRIORITY,-),
=(OPRIORITY,"1-URGENT") 
>(OTOTALPRICE,100000.00).
	\end{lstlisting}%
	\begin{lstlisting}[language=Java,caption=Message translation program: $mt_{tip}$,label={lst:mapping},numbers=left,numbersep=1pt]
conv-order(id,otype,
ORDERKEY,CUSTKEY,SHIPPRIORITY):-
order(id,otype,ORDERKEY,
CUSTKEY,-,SHIPPRIORITY,-).
	\end{lstlisting}%
\end{minipage}%
\hfill
\begin{minipage}[t]{.45\textwidth}
	\begin{lstlisting}[language=Java,caption=Routing condition with \texttt{join} over ``multi-format" message,label={lst:join},numbers=left,numbersep=1pt]
cbr-cust(CUSTKEY,-):-
customer(cid,ctype,CUSTKEY,-,
CNATIONKEY,-,ACCTBAL,-), 
nation(nid,ntype,NATIONKEY,-,
NREGIONKEY,-),
>(ACCTBAL,3000.0),
=(CNATIONKEY,NATIONKEY)
=(NREGIONKEY,3).
	\end{lstlisting}%
\end{minipage} %

\begin{table}
	\caption{Throughput measurements for format conversion, message routing and tranformation patterns based on $4$kB messages generated from 1.5 million standard TPC-H orders records.}
	\label{tab:results}
\begin{tabular}{lcccccccc}
	\multirow{3}{*}{} &
	\multicolumn{1}{c}{\textbf{Format}} &
	\multicolumn{4}{c}{\textbf{Content-based Routing}} &
	\multicolumn{2}{c}{\textbf{Message Transformation}} \\
	\hline
	& \parbox[t]{1cm}{Conv. (msec)} & \parbox[t]{1cm}{Time (sec)} & \parbox[t]{1cm}{Mean (tps)} & \parbox[t]{1cm}{Time (sec)} & \parbox[t]{1cm}{Mean (tps) (Join)} & \parbox[t]{1cm}{Time (sec)} & \parbox[t]{1cm}{Mean (tps)} \\
			\hline \hline
			\parbox[t]{1cm}{\textbf{Camel-Datalog}} & \parbox[t]{1.5cm}{7,239.60 +/-152.69} & \parbox[t]{0.3cm}{\textbf{108}} & \parbox[t]{1.5cm}{\textbf{13,761.47} +/-340.08} & \parbox[t]{0.3cm}{126} & \parbox[t]{1.5cm}{11,904.76 +/-261.20} & \parbox[t]{0.3cm}{\textbf{103}} & \parbox[t]{1.5cm}{\textbf{14,423.08} +/-228.74} \\
			\parbox[t]{1.5cm}{\textbf{Camel-Java}} & \parbox[t]{1.5cm}{\textbf{6,648.50} +/-143.55} & \parbox[t]{0.3cm}{115} & \parbox[t]{1.5cm}{12,931.03 +/-304,90} & \parbox[t]{0.3cm}{\textbf{117}} & \parbox[t]{1.5cm}{\textbf{12,633.26} +/-176.89} & \parbox[t]{0.3cm}{107} & \parbox[t]{1.5cm}{13,888.89 +/-247.40} \\
			\hline
			\parbox[t]{1.5cm}{\textbf{Datalog-Bulk} (size=10)} & \parbox[t]{1.5cm}{7,239.60 +/-152.69} & \parbox[t]{0.3cm}{12} & \parbox[t]{1.5cm}{122,467.58 +/-2,532.42} & \parbox[t]{0.3cm}{13} & \parbox[t]{1.5cm}{116,780.00 +/-1714,92} & \parbox[t]{0.3cm}{11} & \parbox[t]{1.5cm}{133,053.20 +/-1,645.39} \\
		\end{tabular}	
\end{table}	

The throughput test streams all $1.5$ million order / customer messages to the pipeline. The performance measurement results are depicted in Table \ref{tab:results} for a single thread execution. Measurements with multiple threads show a scaling up to factor $10$ of the results, with a saturation around $36$ threads (\ie factor of number of cores; not shown). The stream conversion to JSON object aggregated for all messages is slightly faster than for ONC. However, in both order messages cases the TIP-based implementation reaches a slightly higher transaction per second rate (tps), which lets the processing end 7 s and 4 s earlier respectively, due to the natural processing of ONC iterators in the Datalog engine. Although the measured 99\% confidence intervals do not overlap, the execution times are similar. The rather theoretical case of increasing the number of selection / built-in operations on the order messages (\eg date before / after, string contains) showed a stronger impact for the Camel-Java case than the Camel-Datalog case (not shown). 
In general, the Camel-Java implementation concludes with a routing decision as soon as a logical conjunction is found, while the conjunctive Datalog implementation currently evaluates all conditions before returning. In the context of integration operations this is not necessary, thus could be improved by adapting the Datalog evaluation for that, which we could experimentally show (not shown; out of scope for this paper). The measured throughput of the content-based router with \texttt{join} processing on \enquote{multi-format} the 1.5 million TPC-H customer / nation messages again shows similar results. Only this time, the too simple \texttt{NestedLoopJoin} implementation in the used Datalog engine causes a loss of 9 seconds compared to the \enquote{hand-coded} JSON join implementation.

	
\subsection{Outlook: \enquote{Multi-record} Table Message Processing}
The discussed measurements assume that a message has a \enquote{single-record} payload, which results in $1.5$ million messages with one record / message identifier each. So far, the JSON to ONC conversion creates ONC collections with only one table iterator (to conform with EIP semantics). However, the nature of our approach allows us to send ONC collections with several entries (each representing a unique message payload with message identifier). Knowing that this would change the semantics of several patterns (\eg the content-based router), we conducted the same test as before with \enquote{multi-record} table messages of bulk size 10, which reduces the required runtime to 12 s for the router and 11 s for the translator, which can still be used with its original definition (cf. Table \ref{tab:results}). Increasing the bulk size to 100 or even $1,000$ reduces the required time to 1 s, which means that all $1.5$ million messages can be processed with one step in one single thread. Hereby, increasing the bulk size means reducing the number of message collections, while increasing the rows in the single collection. The impressive numbers are due to the efficient table-centric Datalog evaluation on fewer, multi-row message collections. The higher throughput comes with the cost of a higher latency. The noticed join performance issue can be seen in the Datalog-bulk case as well, which required $13$ steps / seconds to process the $1.5$ million customer / nation messages.
	
\section{Related Work}\label{sec:relatedwork}
The application of table-centric operators to 
current integration systems has not been considered before, up to our knowledge, and was only recently introduced by our position paper~\cite{DBLP:conf/dexa/RitterB14}, which discusses the expressiveness of table-centric / logic programming for integration processing on the content level.
	
The work on Java systems like Telegraph Dataflow \cite{DBLP:journals/sigmod/ShahMFH01}, and Jaguar \cite{DBLP:journals/concurrency/WelshC00}) can be considered related work in the area of programming languages on application systems for faster, data-aware processing. These approaches are mainly targeting to make Java more capable of data-processing, while mainly dealing with threading, garbage collection and memory management. None of them considers the combination of the host language with table-centric processing.
	
\paragraph{Declarative XML Processing} Related work can be found in the area of declarative XML message processing (\eg \cite{DBLP:conf/cidr/0002KM07}). Using an XQuery data store for defining persistent message queues (\ie conflicting with \emph{O3}), the work targets a complementary subset of our approach (\ie persistent message queuing).
	
	
\paragraph{Data Integration} The data integration domain uses integration systems for querying remote data that is treated as local or \enquote{virtual} relations. Starting with SQL-based approaches (\eg using \texttt{Garlic}~\cite{DBLP:conf/vldb/HaasKWY97}), the data integration research reached relational logic programming, summarized by \cite{DBLP:series/synthesis/2010Genesereth}. In contrast to such remote queries, 
we define a table-centric, integration programming approach for application integration, while keeping the current 
semantics (for now).
	
\paragraph{Data-Intensive and Scientific Workflow Management} Based on the data patterns in workflow systems described by Russel et al.\ \cite{DBLP:conf/er/RussellHEA05}, modeling and data access approaches have been studied (\eg by Reimann et al.\ \cite{DBLP:conf/btw/ReimannS13}) in simulation workflows. The basic data management patterns in simulation workflows are ETL operations (\eg format conversions, filters), a subset of the EIP and can be represented among others by our approach. The (map/reduce-style) data iteration pattern can be represented by combined EIPs like \emph{scatter/gather} or \emph{splitter/gather}.
	
Similar to our approach, data and control flow have been considered in scientific workflow management systems~\cite{DBLP:conf/vldb/VrhovnikSSMMMK07},
which run the integration system optimally synchronized with the database. However, the work exclusively focuses on the optimization of workflow execution, not integration systems, and does not consider the usage of table-centric programming on the application server level.
	
\section{Concluding Remarks} \label{sec:discussion}
This paper motivates a look into a growing \enquote{processing discrepancy} (\eg message throughput) between current integration and complementary systems (\eg MQ, RDBMS) based on known scenarios with new requirements and fast growing new domains (\emph{O1--O3}). Towards a message throughput improvement, we extended the current integration processing on a content level by table-centric integration processing (TIP). To remain compliant to the current EIP definitions the TIP-operators work on \enquote{single-record} messages, which lets us compare with current approaches using  a brief experimental throughput evaluation. Although the results slightly improve the standard processing, not to mention the declarative vs. \enquote{hand-coded} definition of integration content, the actual potential of our approach lies in \enquote{multi-record} table message processing. However, that requires an adaption of some pattern EIP definitions, which is out of scope for this paper.
	
	
\paragraph{Open Research Challenges} For a more comprehensive, experimental evaluation, an EIP micro-benchmark will be developed on an extension of the TPC-H and TPC-C benchmarks. EIP definitions do not discuss streaming patterns / operators, which could be evaluated (complementarily) based on Datalog streaming theory (\eg \cite{DBLP:conf/sebd/Zaniolo12,DBLP:conf/datalog/Zaniolo12}). Eventually, the existing EIP definitions have to be adapted to that and probably new patterns will be established. Notably, the used Datalog engine has to be improved (\eg join evaluation) and enhanced for integration processing (\eg for early-match / stop during routing).

	
\section*{Acknowledgements}
We especially thank Dr.\ Fredrik Nordvall Forsberg and Prof. Dr. Erhard Rahm for proof-reading and valuable discussions on the paper.

\bibliographystyle{splncs03}
\bibliography{datalogblocks}

\newpage
\appendix
\section{Supplementary Material}
In this section supplementary material is listed and briefly explained. The material provides further details on the described table-operator content / Datalog rules used for the experimental evaluation and can be seen as examples for the TIP definitions. In addition, the Apache Camel routes and the results of the experimental evaluation are shared for comparison and traceability.

The supplementary material is provided inline for better readability and can be externalized to a separate document, if required.

\subsection{Details on the Test Setup} \label{supp:setupDetails}
During the execution of our micro-benchmark in Java, we used the \emph{JMH} code tool from \emph{OpenJDK}\footnote{OpenJDK JMH, visited 02/2015: \url{http://openjdk.java.net/projects/code-tools/jmh/}.} to overcome the \enquote{profile-guided optimization}.

\paragraph{Data and Message Generation}
The message generation is a \enquote{two-step} approach consisting of a data generation and a message generation step. The data generation step uses the standard TPC-H generator on scale level one, from which we took the generated order records (\ie CSV format), converted them to the JSON format, added additional message identifier and type fields, and stored them in one big JSON array. Each JSON object represents the payload of a single message (\ie \enquote{single-record} message). During the test preparation, Camel message payloads are constructed as streams of these JSON objects and sent into the pipeline. In each individual test, the processors in the pipeline have to convert the JSON stream using the (Stax-like) Jackson Stream API\footnote{Jackson Stream API, visited 02/2015: \url{http://wiki.fasterxml.com/JacksonStreamingApi}.} to either a Java or ONC represenation (cf. canonical data model; required an extension of Jackson for ONC), before applying the specfic pattern to test. That means, the conversions will be part of the measurements, however, extracted to a separate performance figure. We have decided to take the TPC-H data generator due to its applicability to our routing and transformation tests and since it is well-known in our domain.

\paragraph{Message Routing and Transformation}
The selection / built-in operations are mainly used during message routing (\eg content-based router, message filter). For the evaluation, we have defined a routing condition (cf. List. \ref{lst:cbr}), which routes the message to its receiver, if and only if, two conditions are fulfilled for routing condition $tip\_{rc}$: \emph{conds} = $\{$string equality, integer less than\} on fields \emph{body.x} = \{objecttype, ototalprice\}. The $help\_rc$ function is implemented in Java to pass the expected value to the runtime system on system-level. The corresponding pure-Java, content-level implementation uses JSON path statements for $O(1)$ element access and conducts the type-specific condition evaluation \enquote{hand-coded}. The routing condition is defined to route $904,567$ of the $1.5$ million messages. The projection operation is mostly used in message tranformation patterns (\eg message translator, content modifier), from which we selected a message translator. The translator projects the fields of the incoming message to a target format (cf. List. \ref{lst:mapping}) using our $mt_{tip}$ implementation or a \enquote{hand-coded} projection on the Java Object representation. One requirement from the mentioned scenarios (\emph{O1}) are multi-format, multi-record messages. This type of messages has one \enquote{primary} message body and (several) dependent sub-entries usually in a different format. The sub-entries act as \enquote{temporary} processing context (\eg added by a content enricher), which can be pulled off, before forwarding the message to its receiver. For the evaluation, we use the join together with selection / built-in operations for a content-based router on a TPC-H customer message as JSON object embedded in a JSON array with all entries of the nation table. The message is routed, if and only if, the customers actual balance (\texttt{ACCTBAL}) is bigger than $3,000$ and the customer is from the European region determined through join via nation key (cf. Suppl. Material in Sect. \ref{app:A}).

\subsection{Datalog Rules} \label{app:A}
The Datalog rules, used for the table-centric integration processing on the TPC-H order, customer and nation messages during the experimental analysis are listed subsequently.

Listing \ref{lst:cbr} denotes the rule for the content-based routing pattern on a TPC-H order message, which is used to route orders to the first recipient, if and only if, the order is urgent (\texttt{OPRIORITY}=='\texttt{1-URGENT}') and costly (\texttt{OTOTALPRICE}$>100,000.00$). In the message translation evaluation, order messages are translated to a target format with only the primary and foreign key fields and the shipment priority according to the rule in List. \ref{lst:mapping}. The extended routing condition in List. \ref{lst:join} is used to route TPC-H customer record messages with the complete list of TPC-H nation records, if and only if, the actual balance is above average (\texttt{ACCTBAL}$>3,000$) and the customer is from \emph{Europe} region (\texttt{NREGIONKEY}==$3$).

\subsection{Integration Scenarios: Apache Camel Routes} \label{supp:routes}
In this section, the integration scenarios, used for the experimental analysis, are described in more detail.

Listing \ref{lst:camel-data-cbr} and List. \ref{lst:camel-cbr} denote the Camel routes / message channels for the content-based routing cases. The \texttt{direct}-endpoint is used to receive a stream of messages in JSON format, which are then serialized into a Java Object or ONC-iterator form, the canonical data model (CDM). On the CDM subsequent operations can be defined and executed. The Camel \texttt{choice} defines a conditional routing based on a Camel \texttt{expression} evaluating the message's body / data. The expression is one of the extension points in Camel to which our TIP operations can be applied (\eg $rc_{tip}$). In the standard Camel-Java case a \enquote{hand-coded} Java expression, \texttt{new CbrTpchExp()}, is evaluated. If the expressions evaluate to \texttt{false}, the messages are processed in the \texttt{otherwise} channel. The same routes are used for the normal and the \enquote{multi-format} router cases, only the expressions are exchanged.

Another Camel interface that allows the definition of own content-level extensions is the \texttt{processor}. To evaluate the message translation capabilities, we applied our TIP operators to the Camel processor to access and change the ONC messages. In the Camel-Java case a \enquote{hand-coded} processor is implemented as \texttt{new MtTpchProc()}. Both processors receive a message and translate its content, before handing it back to the system-level processing.

\par\noindent
\begin{minipage}[t]{.45\textwidth}
\begin{lstlisting}[language=Java,caption=Camel-Datalog route for routing according to $rc_{tip}$,label={lst:camel-data-cbr},numbers=left,numbersep=1pt]
from("direct:datalog-cbr)
.unmarshal()
.choice().when()
.expression(bool_rc(rc_tip(
<conds>))).process(VOID)
.otherwise().process(VOID);
\end{lstlisting}%
\end{minipage}%
\hfill
\begin{minipage}[t]{.45\textwidth}
\begin{lstlisting}[language=Java,caption=Camel-Java route for routing,label={lst:camel-cbr},numbers=left,numbersep=1pt]
from("direct:java-cbr")
.unmarshal()
.choice().when()
.expression(new CbrTpchExp())
.process(VOID)
.otherwise().process(VOID);
\end{lstlisting}%
\end{minipage}%
\hfill
\begin{minipage}[t]{.45\textwidth}
\begin{lstlisting}[language=Java,caption=Camel-Datalog message translation executing a $mt_{tip}$ program,label={lst:camel-mt1},numbers=left,numbersep=1pt]
from("direct:datalog-mt")
.unmarshal()
.process(mt_tip(<program>))
.process(VOID);
\end{lstlisting}%
\end{minipage}%
\hfill
\begin{minipage}[t]{.45\textwidth}
\begin{lstlisting}[language=Java,caption=Camel-Java message translation,label={lst:camel-mt2},numbers=left,numbersep=1pt]
from("direct:java-mt")
.unmarshal()
.process(new MtTpchProc())
.process(VOID);
\end{lstlisting}%
\end{minipage}%

\subsection{Experimental Results} \label{supp:results}
In this section the plotted results of the experimental analysis are shown. Since all tests are based on $1.5$ million TPC-H order or customer / nation (\enquote{multi-format}) messages, the required processing time is represented by the x-axis in steps of one second each. Hence, the earlier the samples stop, the faster the average throughput of the route, which is depicted as y-axis. All measurements were conducted with a single Camel route thread. The first three plots show the two content router and one message translator cases. The fourth one plots the results of the three cases with \enquote{multi-record} table message processing of bulk size of ten. Hence each the depicted discrete throughput measures actually do not represent messages per second, but collections of ten messages each per second (\ie multiplied by factor x10).

\begin{figure}[!ht]
	\begin{center}$
		\begin{array}{cc}
		\subfigure[CBR: Normal]{\label{fig:cbr}\includegraphics[width=0.5\textwidth]{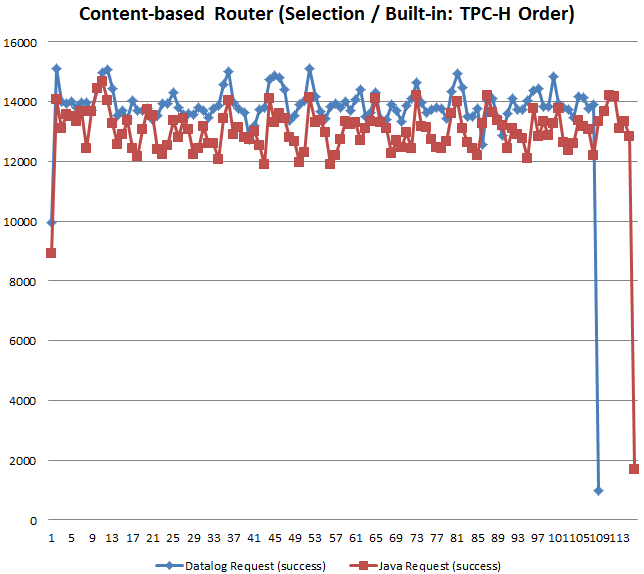}} &
    	\subfigure[CBR: Multi-Msg.]{\label{fig:cbr-join}\includegraphics[width=0.5\textwidth]{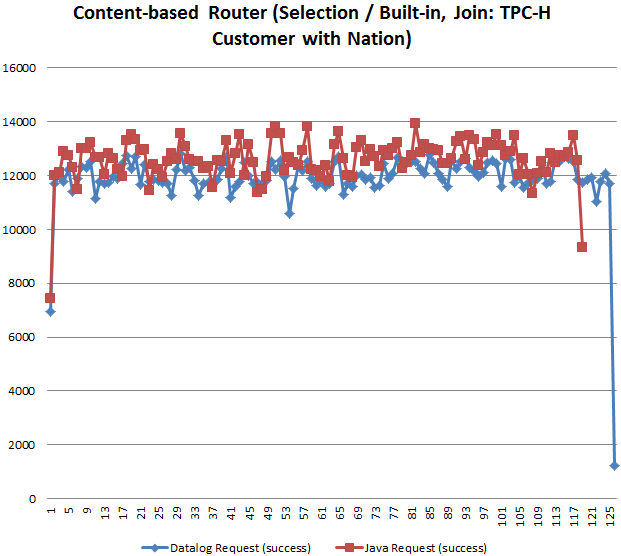}} \\ 		\subfigure[Msg. Translator]{\label{fig:mt}\includegraphics[width=0.5\textwidth]{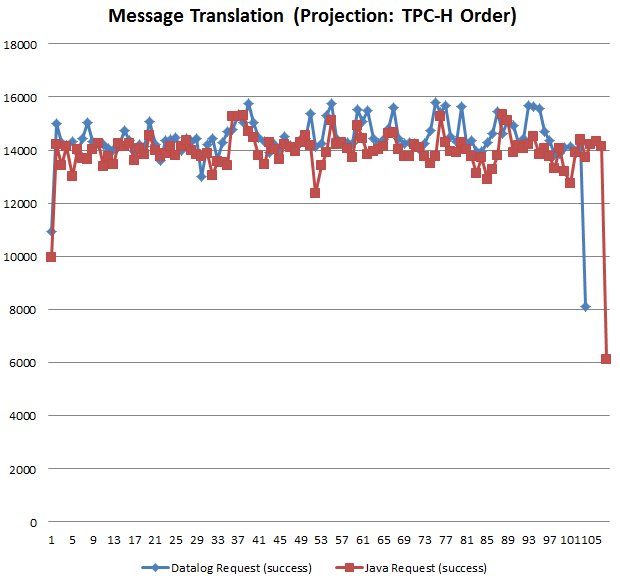}} &
		\subfigure[Table-Processing; x10 msgs/s]{\label{fig:table-b10}\includegraphics[width=0.5\textwidth]{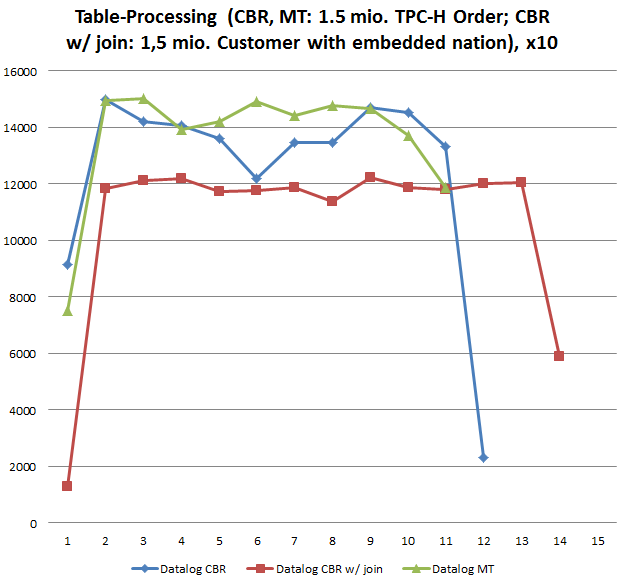}}
		\end{array}$
	\end{center} 
	\caption{Plotted results of the experimental analysis.}
	\label{fig:patterns}
\end{figure}

The results of the first three experiments show that the TIP approach is competitive to \enquote{state-of-the-art} message processing in integration systems. The routing conditions with selection / built operations are better and the nested-loop join implementation is slightly worse than the \enquote{hand-coded} and optimized Java pendent. Hence, an enhanced join implementation will significantly improve the results of the Camel-Datalog evaluation in Fig. \ref{fig:cbr-join}.

\end{document}